\documentclass[prb,twocolumn,showpacs,preprintnumbers,amsmath,amssymb]{revtex4}

\usepackage{graphicx}
\usepackage{dcolumn}
\usepackage{epsfig}

\begin{document}

\title{Normal State Nernst Effect in Electron-doped Pr$_{2-x}$Ce$_x$CuO$_{4-\delta}$: Superconducting Fluctuations and Two-band Transport}

\author{Pengcheng Li}
 \email{pcli@physics.umd.edu}
\author{R. L. Greene}
\affiliation{Center for Superconductivity Research and Department
of Physics, University of Maryland, College Park, Maryland
20742-4111, USA}

\date{\today}

\begin{abstract}

We report a systematic study of normal state Nernst effect in the
electron-doped cuprates Pr$_{2-x}$Ce$_x$CuO$_{4-\delta}$ over a
wide range of doping (0.05$\leq x \leq$0.21) and temperature. At
low temperatures, we observed a notable vortex Nernst signal above
T$_c$ in the underdoped films, but no such normal state vortex
Nernst signal is found in the overdoped region. The
superconducting fluctuations in the underdoped region are most
likely incoherent phase fluctuations as found in hole-doped
cuprates. At high temperatures, a large normal state Nernst signal
is found at dopings from slightly underdoped to highly overdoped.
Combined with normal state thermoelectric power, Hall effect and
magnetoresistance measurements, the large Nernst effect is
compatible with two-band model. For the highly overdoped films,
the large Nernst effect is anomalous and not explainable with a
simple hole-like Fermi surface seen in photoemission experiments.

\end{abstract}


\maketitle

\section{\label{sec:level1}Introduction}

The anomalously large Nernst voltage well above the zero-field
T$_c$ in hole-doped cuprate superconductors is now a well
established experimental observation with a dominant view that it
is due to vortex-like excitations above T$_c$.\cite{WangYY2, Xu}
The appearance of the Nernst signal on approaching T$_c$ from
above marks the onset of a phase uncorrelated pairing amplitude.
The observation of an enhanced diamagnetism near the onset
temperature T$_\nu$ of the anomalous Nernst signal in some
hole-doped cuprates strongly supports the vortex-like excitations
scenario.\cite{WangYY2} The fact that the regime of this large
Nernst effect overlaps with the temperature range where a
pseudogap is seen in the electronic excitation spectrum, also
suggests that the anomalous Nernst effect is related to the
pseudogap phenomenon.

Inspired by the unusual Nernst effect in hole-doped cuprates, many
theories to explain these striking observations have been
proposed. Two of these theories deal with amplitude and phase
fluctuations of the superconducting order parameter. Ussishkin
\textit{et al.}~\cite{Ussishkin} suggest that Gaussian amplitude
fluctuations above T$_c$ are responsible for the Nernst effect for
the optimally-doped and overdoped regimes. For the underdoped
regime they suggest that strong non-Gaussian amplitude
fluctuations can explain the wide temperature range of the
anomalous Nernst signal. The phase fluctuation explanations are
based upon the influential work of Emery and
Kivelson,~\cite{Emery} which preceded the Nernst effect
measurements in underdoped cuprates, and the follow-up work of
Carlson \textit{et al.}~\cite{Carlson} In this theory, the
importance of phase fluctuations is determined by the superfluid
density, $\rho_s$. The smaller the superfluid density, the more
significant the phase fluctuations.~\cite{Emery} In conventional
superconductors, the superfluid density $\rho_s$ is very
large.~\cite{Tinkham, Emery} Phase rigidity, or the strength of
the phase coherence, is so strong that pairing and long-range
order phase coherence occur simultaneously at the transition
temperature T$_c$. The phase degree of freedom plays an
insignificant role in determining the transition temperature and
other relevant properties. However, in the high-T$_c$
superconductors, with a small superfluid density, the long-range
phase coherence is destroyed at T$_c$ while the local Cooper
pairing amplitude remains sizable.

Besides the fluctuation theories, many other theories also
provided possible explanations for the anomalous Nernst effect in
hole-doped cuprates.~\cite{Kontani, Tan, Lee, Ikeda, Alexandrov,
Anderson, Sachdev, Tesanovic} However, none of these explanations
have gained general acceptance and they are still under debate.

Electron-doped cuprates, RE$_{2-x}$Ce$_x$CuO$_{4-\delta}$ (RE=Nd,
Pr, Sm), on the other hand, have demonstrated distinctive results.
Prior Nernst effect experiments~\cite{Fournier,Jiang, Gollink} in
the electron-doped cuprates near optimal doping suggested that
electron-doped cuprates are more conventional than their
hole-doped counterparts, and the superconducting fluctuations are
much weaker, i.e., almost no vortex-like Nernst signal was
observed above T$_c$. In the first part of this paper, a careful
study of the vortex Nernst effect in the electron-doped cuprate
system Pr$_{2-x}$Ce$_x$CuO$_{4-\delta}$ (PCCO) films is reported
and the superconducting fluctuation contribution is reexamined. In
the underdoped region, we observed a vortex Nernst signal well
above T$_c$, but no such normal state vortex Nernst signal is
found in the overdoped region. As in the hole-doped cuprates, we
found the stronger superconducting fluctuations in the underdoped
PCCO are also compatible with idea of incoherent phase
fluctuations.

In the normal state, prior Nernst effect on oxygen-doped
Nd$_{1.85}$Ce$_{0.15}$CuO$_{4-\delta}$ (NCCO) observed a large
Nernst signal,~\cite{Jiang,Fournier,Gollink} which has a
distinctively different temperature and field dependence from the
vortex Nernst signal found in the hole-doped cuprates. The sign
change of the Hall coefficient and the enhanced normal state
Nernst effect were argued to result from a two-carrier (electron
and hole) quasi-particle contribution around optimal doping of
Ce=0.15. This intriguing two-band behavior observed in transport
experiments was later confirmed by angle resolved photoemission
spectroscopy (ARPES).~\cite{Armitage,Matsui} These measurements
found that the Fermi surface (FS) evolves from a Mott insulator
parent compound to an electron-like FS centered at ($\pi$,0) in
the underdoped region. At optimal doping, a hole-like FS pocket
centered at ($\pi/2$, $\pi/2$) coexists with an electron-like
pocket near ($\pi$, 0)and (0, $\pi$). We mention that low
temperature penetration depth measurements in PCCO are also
consistent with a weakly-coupled two-band model~\cite{Xiang}. From
the evolution of band structure with doping, one expects a
hole-like FS centered at ($\pi,\pi$) in the overdoped region. In
fact, Matsui \emph{et al.}~\cite{Matsui2} recently observed a
large hole-like pocket in an overdoped
Nd$_{1.83}$Ce$_{0.17}$CuO$_4$.

Most of the previous transport measurements were performed on
optimally-doped Nd$_{1.85}$Ce$_{0.15}$CuO$_{4-\delta}$ and the
charge doping was varied by oxygen content. The evolution of these
transport properties with oxygen reduction suggested that the hole
band is controlled by the oxygen content: a single electron band
in the fully oxygenated regime to a two-band regime in the
superconducting phase and a hole-like band in the deoxygenated
state. Recently, Balci \textit{et al.}~\cite{HamzaNernst} observed
a large normal state Nernst signal in PCCO films with Ce
concentration varied around optimal-doping. This is consistent
with previous results in the oxygen-doped NCCO. However, more
detailed studies of the Nernst effect in electron-doped cuprates
over a wide range of Ce concentration are lacking and it is
important to investigate the transport properties in the very
underdoped or overdoped regimes since useful information could be
obtained for further understanding of the band structure (and
scattering) at the extreme dopings. In the second part of this
paper, we report our extensive normal state Nernst effect
measurements on Pr$_{2-x}$Ce$_x$CuO$_{4-\delta}$ over a wide range
of doping (\textit{x}=0.05 to 0.21). We found that the normal
state Nernst signal is large around the optimal doping, in
agreement with previous reports. In the slightly underdoped and
highly overdoped films \textit{x}=0.11 and 0.19, the Nernst signal
is still large. This is contrary to what is expected for a simple
single carrier system, suggesting that the transport in this
regime may be influenced by anomalous energy dependent scattering
at the Fermi surface. For the extremely underdoped
\textit{x}=0.05, the Nernst signal decreases rapidly, consistent
with a simple single carrier system at this doping.

\section{Experiments details}

High quality PCCO films with thickness about 2500-3000 \AA\ were
fabricated by pulsed laser deposition on SrTiO$_3$ substrates
(10$\times$5 mm$^2$).~\cite{Maiser, Peng} The films were
characterized by AC susceptibility, resistivity measurements and
Rutherford Back Scattering. The minimum channelling yield obtained
was 10\% to 20\% indicating a good epitaxial growth. A sharp
transition ($\triangle T_c<$1 K) indicates that our films are of
high quality. We note that since the oxygen content has an
influence on both the superconducting and normal state properties
of the material,~\cite{Jiang} we optimized the annealing process
for each Ce concentration as in Ref.~\cite{YoramQCP}. The sharp
transition, low residual resistivity and the Hall coefficient are
exactly the same as the previous report.~\cite{YoramQCP} Since the
exact content of oxygen cannot be determined in films, we use the
low temperature values of the Hall coefficient and Ce content to
determine the temperature versus doping phase diagram. The films
of size of 10$\times$5 mm$^2$ used in Nernst effect experiments
were patterned into a standard Hall bar by ion-mill technique.

The Nernst measurements were performed using a
one-heater-two-thermometer technique. The film was attached on one
end to a copper block with a mechanical clamp and the other end
was left free. A small chip resistor heater is attached on the
free end, and a temperature gradient is created by applying a
constant current to the heater. Two tiny Lakeshore Cernox
thermometers are attached on the two ends of the sample to monitor
the temperature gradient continuously. The Nernst voltage is
measured with a Keithley 2001 multimeter with a 1801 preamp while
the field is slowly ramped at a rate of 0.3 T/min between -9 T and
+9 T ($H\perp ab$). The system temperature was well controlled to
give stability of the temperature of $\pm$1 mK, which enables us
to perform a high resolution Nernst voltage measurement (typically
$\sim$10 nV in our setup). The temperature gradient is around
0.5-2 K/cm depending on the temperature of measurement, and the
sample temperature is taken as the average of hot and cold end
temperatures. The Nernst signal is obtained by subtracting
negative field data from positive field data to eliminate any
possible thermopower contribution. The Nernst signal was defined
as
\begin{equation}\label{1a}
e_y=\frac{E_y}{-\nabla T}
\end{equation}
where $E_y$ is the transverse electrical field across the sample
and $-\nabla T$ is the temperature gradient along its length.

\section{Vortex Nernst effect in $Pr_{2-x}Ce_xCuO_{4-\delta}$}

\subsection{Experimental results}

It is known that in a type II superconductor below T$_c$, vortices
moving under a longitudinal thermal-gradient and perpendicular
magnetic field will generate a transverse electric field, known as
the vortex Nernst signal and the sign of the vortex Nernst signal
is usually defined as positive.~\cite{OngNP} The vortex Nernst
signal is usually large compared to the normal state signal. When
the temperature is far above T$_c$, quasi-particles are the only
source for a Nernst signal. For a one-carrier system the normal
state Nernst effect is much smaller than the vortex Nernst
effect.~\cite{WangYY2}

\begin{figure}
\begin{center}
\includegraphics[scale=1]{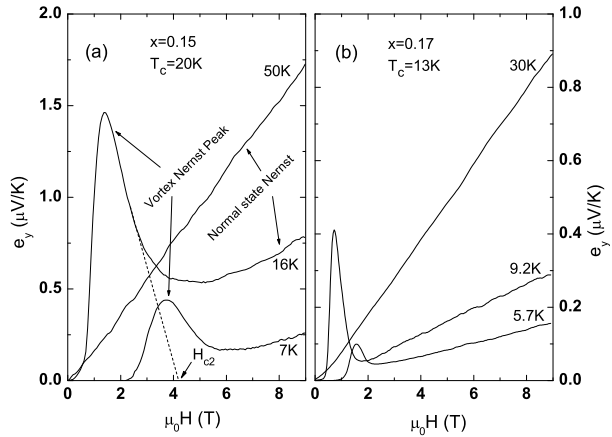}
\end{center}
\caption{Nernst signal as a function of magnetic field in an
optimally-doped \textit{x}=0.15 (a) and an overdoped
\textit{x}=0.17 (b) PCCO films at different temperatures.}
\label{Fig1}
\end{figure}

Figure~\ref{Fig1} illustrates the vortex Nernst effect and the
normal state Nernst signal in the optimally-doped and overdoped
PCCO films. Below T$_c$ and H$_{c2}$, a sharp and large vortex
Nernst peak is seen, which starts from the melting field (the
magnetic field at which the vortex is depinned). Above H$_{c2}$,
the Nernst signal is linear in field, as also found when the
temperature is above T$_c$. This linear field dependent Nernst
signal is attributed to the normal state quasi-particles. In the
figure, we also show a normal state Nernst curve for T$>$T$_c$
(T=50 K for \emph{x}=0.15 and 30 K for \emph{x}=0.17). We find
that the normal state Nernst signal at higher field is even larger
than the peak in vortex Nernst signal below T$_c$. This is in
striking contrast to hole-doped cuprates, in which the
quasi-particle Nernst signal is much smaller than the vortex
Nernst signal.~\cite{WangYY2} To obtain the net vortex Nernst
signal, the linear normal state Nernst signal can be subtracted
from the measured data.

\begin{figure}
\begin{center}
\includegraphics[scale=0.8]{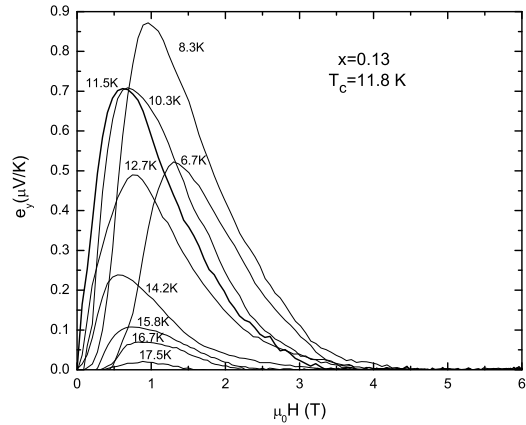}
\end{center}
\caption{Vortex Nernst effect in an underdoped \emph{x}=0.13 PCCO
film at different temperatures after subtraction of the linear
normal state background. Thicker line is the Nernst signal at
T$\approx$T$_c$.} \label{Fig2}
\end{figure}

We carefully studied the Nernst effect around T$_c$ to search for
possible superconducting fluctuation effects in PCCO, especially
in the underdoped regime. Fig.~\ref{Fig2} shows the low
temperature vortex Nernst effect result for an underdoped film
\textit{x}=0.13 (T$_c$=11.8 K, from the peak temperature of the
imaginary part of susceptibility) after subtraction of the linear
normal state Nernst signal. The peak-featured vortex Nernst signal
is observed to persist to temperatures higher than T$_c$. As seen
in Fig.~\ref{Fig2}, the vortex Nernst signal is still robust at
T=17.5 K, which is about 6 K higher than T$_c$. When temperature
is above T$_c$ (T$>$20 K for \textit{x}=0.13), the vortex Nernst
signal vanishes and the linear quasi-particle field dependent
Nernst signal is recovered. The Nernst effect measurement was also
performed on PCCO films with dopings \textit{x}=0.14, 0.15, 0.16,
0.17, 0.19. For \textit{x}=0.14 and \textit{x}=0.15, the result is
similar to \textit{x}=0.13, but the onset temperature of the
vortex Nernst signal (temperature where a vortex Nernst peak
appears) is about 4 K above T$_c$ for \textit{x}=0.14 and 3 K
above T$_c$ for \textit{x}=0.15. However, in the overdoped films,
the vortex Nernst signal (low field peak) vanishes immediately at
the superconducting transition temperature T$_c$. Moreover, the
linear quasi-particle Nernst signal emerges when temperature is
just above T$_c$, suggesting that the normal state is recovered,
with minimal superconducting fluctuations, right at T$_c$.

\begin{figure}
\begin{center}
\includegraphics[scale=0.6]{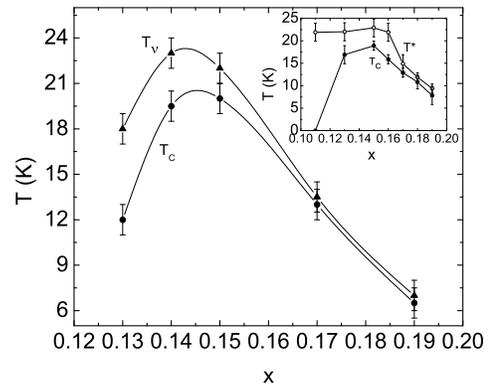}
\end{center}
\caption{Doping dependence of T$_c$ and the onset temperature
T$_\nu$ of vortex Nernst signal. Inset shows recent planar
tunneling data in PCCO,~\cite{YoramTunneling} T$^*$ is the
temperature where a normal state gap opens. }\label{Fig3}
\end{figure}

Figure~\ref{Fig3} displays the onset temperature of the vortex
Nernst signal along with the superconducting transition
temperature T$_c$ as a function of doping for PCCO. It is clear
that in the underdoped region the difference between these two
characteristic temperatures is larger than the overdoped side,
suggesting a larger superconducting fluctuation effect in the
underdoped region with a more conventional behavior in the
overdoped region.

\subsection{Discussion}

How do we understand the superconducting fluctuations in the
electron-doped cuprates?  It is known that superconductivity is
characterized by a complex order parameter $|\psi|exp(-i\theta)$,
with an amplitude $|\psi|$ and a phase $\theta$ at each space
point.~\cite{Tinkham} Fluctuations in either amplitude or phase
will affect the superconducting properties. The conventional
fluctuation theories primarily deal with thermal fluctuations of
the amplitude $|\psi|$ of the order parameter.~\cite{Tinkham,
Larkin} By solving the time-dependent Ginzburg-Landau equation in
the Gaussian approximation, Ussushkin \textit{et
al.}~\cite{Ussishkin} calculated the transverse thermoelectric
coefficient which results from the amplitude fluctuations:
\begin{equation}\label{1}
\alpha^{SC}_{xy}=\frac{1}{6}\frac{e}{\hbar}\frac{\xi^2}{l^2_B}\propto
\frac{1}{T-T_c}
\end{equation}
where $\alpha^{SC}_{xy}$ is the transverse Peltier conductivity,
$l_B=(h/eB)^{1/2}$ is the magnetic length and $\xi$ is the
in-plane coherence length.

The vortex Nernst signal, $e_y=\alpha^{SC}_{xy}/\sigma_{xx}$
($\sigma_{xx}$ is the conductivity), is mainly proportional to the
in-plane coherence length $\xi_{ab}$. In the electron-doped
cuprates, the in-plane coherence length increases with doping
(H$_{c2}$ decreases rapidly with doping~\cite{FournierHc2}) and
thus from Eq.~\ref{1} one would expect a stronger fluctuation
effect as doping increases. However, the absence of fluctuation
effects in the overdoped PCCO contradicts this theoretical
expectation. In addition, compared to the hole-doped cuprates
(e.g. LSCO), electron-doped cuprates have a much longer coherence
length (about one order of magnitude larger~\cite{FournierHc2,
HamzaNernst}), yet a weaker fluctuation effect is observed: the
fluctuation regime, $\triangle T_{fl}=T_{\nu}/T_c -1$ is about 0.5
for PCCO with \textit{x}=0.13 and 4 for LSCO with \textit{x}=0.1
(T$_c$=20 K).~\cite{WangYY2} These two results strongly suggest
that conventional amplitude fluctuation theory can not explain the
Nernst effect results in underdoped electron-doped cuprates.

Now we turn to phase fluctuations, which are claimed to explain
the anomalous Nernst effect in the hole-doped cuprates. In
conventional superconductors, phase fluctuations plays an
insignificant role in determining the transition temperature
because of the large superfluid density $\rho_s$. However, in the
high-T$_c$ superconductors, the proximity to the Mott insulator
leads to a very small superfluid density. Therefore, it is
possible that long-range phase coherence is destroyed at T$_c$
while the local Cooper pairing amplitude remains sizable. The
significance of the phase fluctuation can be assessed by
evaluating the phase stiffness temperature,~\cite{Emery}
$T^{max}_{\theta}=A\rho_s(0)d/m^*$, at which phase order would
disappear (here $d$ the spacing between adjacent CuO$_2$ layers,
$A$=0.9 is a numeric factor for quasi-2D systems and $m^*$ the
effective mass). If T$_c\ll$T$^{max}_{\theta}$, phase fluctuations
are relatively unimportant, and T$_c$ will be close to the
mean-field transition temperature, T$^{MF}_c$, as predicted by BCS
theory. On the other hand, if T$^{max}_{\theta}\approx$T$_c$, then
T$_c$ is determined primarily by phase ordering, and T$^{MF}_c$ is
simply the characteristic temperature below which pairing becomes
significant locally. In conventional superconductors,
T$^{max}_{\theta}$ is orders of magnitude larger than T$_c$, and
phase coherence is so strong that T$_c$ is the Cooper pair
formation temperature. In the overdoped hole-doped cuprates,
T$^{max}_{\theta}$/T$_c$ is around 2-5, suggesting that phase
fluctuations become more important. In the underdoped hole-doped
cuprates, T$^{max}_{\theta}$ is very close to T$_c$, suggesting
that phase fluctuations are dominant in determining the
superconducting phase transition. The fluctuations suppress the
transition temperature from the mean-field value T$^{MF}_{c}$, at
which Cooper pairs form, to the observed T$_c$ where long-rang
phase coherence is established. In the electron-doped cuprate
PCCO, penetration depth measurements gave $\lambda^{-2}(0)$=9, 15
and 40 $\mu m^{-2}$ for \textit{x}=0.13, 0.15 and 0.17
respectively.~\cite{{Snezhko}} Using the relation
$\rho_s(0)=m^*/e^2\lambda^2(0)$, we can estimate the
T$^{max}_{\theta}$/T$_c$ value for PCCO. Simple calculation gives
T$^{max}_{\theta}$/T$_c$=2, 2.4 and 11 for 0.13, 0.15 and 0.17
respectively. Thus, phase fluctuations are suppressed in the
overdoped region since $\frac{T^{max}_{\theta}}{T_c}\gg 1$ and
then $T_c\approx T^{MF}_c$. Although the values of
T$^{max}_{\theta}$/T$_c$ of the underdoped and optimally-doped
PCCO are close to those of optimally-doped and overdoped LSCO, the
fluctuation regime is somewhat narrower in PCCO than LSCO
($\triangle T_{fl}$=0.5$\pm$0.1 for PCCO with $x$=0.13,
1.2$\pm$0.1 for LSCO $x$=0.17.~\cite{WangYY2}) A likely reason is
the weaker coupling of the Cooper pairs in the electron-doped
cuprates. The pairing amplitude measured by the superconducting
gap is about 3.5 meV for PCCO $x$=0.13, smaller than the gap value
($\sim$10 meV) for the optimally doped LSCO.~\cite{Ino} The value
of $2\triangle/k_BT_c$ is smaller in PCCO ($\sim$ 3.5) than in
LSCO ($\sim$ 8). This suggests that the temperature driven
pair-breaking effect is stronger in PCCO, so that any local
pairing can not survive with increasing temperature. Thus, the
thermal pair breaking competes with the phase fluctuation effect
and reduces the size of the fluctuation region.

\begin{figure}
\begin{center}
\includegraphics[scale=1]{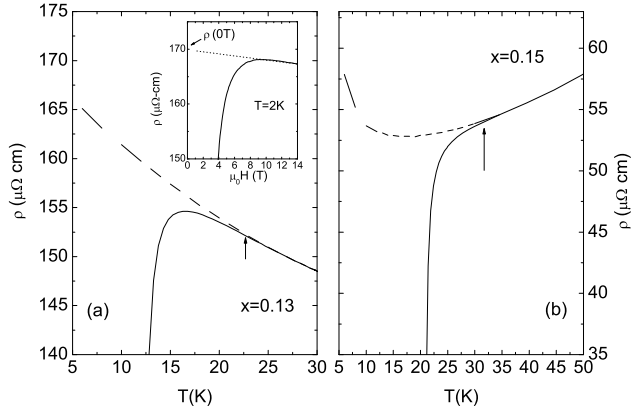}
\end{center}
\caption{Paraconductivity of an underdoped $x$=0.13 (a) and an
optimally-doped $x$=0.15 (b) PCCO films. Solid line is the
zero-field resistivity. Inset shows the procedure of obtaining the
zero-field normal state resistivity [dashed lines in (a) and
(b)].} \label{Fig4}
\end{figure}

In addition, our resistivity measurements on the underdoped and
optimally-doped PCCO films also show some interesting results. As
displayed in Fig.~\ref{Fig4}, the solid curves are the zero-field
resistivity of $x$=0.13 and 0.15. In order to eliminate the
negative magnetoresistance at these dopings, we measured the
in-plane resistivity versus field at different temperatures and
extrapolated the data to get the ``effective" zero-field normal
state resistivity. Its temperature dependence is shown as dashed
curve in Fig.~\ref{Fig4}. The arrows mark the temperature where
the zero-field resistivity deviates from the zero-field normal
state resistivity. Interestingly, the onset temperature of the
deviation is slightly higher than the onset temperatures of the
vortex Nernst effect in both underdoped and optimally-doped
samples. Although the excessive conductivity (paraconductivity) is
usually attributed to amplitude fluctuations,~\cite{Larkin} this
is not certain in PCCO case because phase fluctuations could also
induce the enhanced conductivity above T$_c$. However, to
distinguish these fluctuations is difficult theoretically and
experimentally and it is beyond our knowledge and the scope of
this paper.

We mention that a planar tunneling experiment in PCCO observed a
normal state energy gap throughout the entire doping
range.~\cite{YoramTunneling} This normal state gap persists to a
temperature higher than the superconducting transition temperature
T$_c$ in the underdoped region but follows T$_c$ on the overdoped
side (inset of Fig.~\ref{Fig3}). This was interpreted as a result
of finite pairing amplitude above T$_c$ in the underdoped region.
The onset temperature of the vortex Nernst signal extends into
this region, suggesting that may be related to the normal state
gap seen the tunneling.

\subsection{Conclusion}

To summarize this part, we carefully investigated the
superconducting fluctuation effects in PCCO films by Nernst effect
measurements. We found that the fluctuations are stronger in the
underdoped region than in the overdoped region. We argued against
amplitude fluctuations as an explanation of this observation. It
is likely that phase fluctuations are responsible for the normal
state vortex signal in the underdoped PCCO films. This is
consistent with the anomalous Nernst effect found in the
hole-doped cuprates.

\section{Normal state Nernst effect in $Pr_{2-x}Ce_xCuO_{4-\delta}$}

\subsection{Experimental results}

Electron-doped cuprates are distinct from hole-doped cuprates in
having a low H$_{c2}$ and thus the normal state can be easily
accessed for temperature below T$_c$. As shown before, when the
external magnetic field is larger than H$_{c2}$ or temperature
higher than T$_c$, the Nernst signal $e_y$ in PCCO is linear in
field. The Nernst coefficient $\nu$ is defined as the slope of
$e_y(B)$, i.e., $\nu\equiv e_y/B$.

\begin{figure}
\begin{center}
\includegraphics[scale=1]{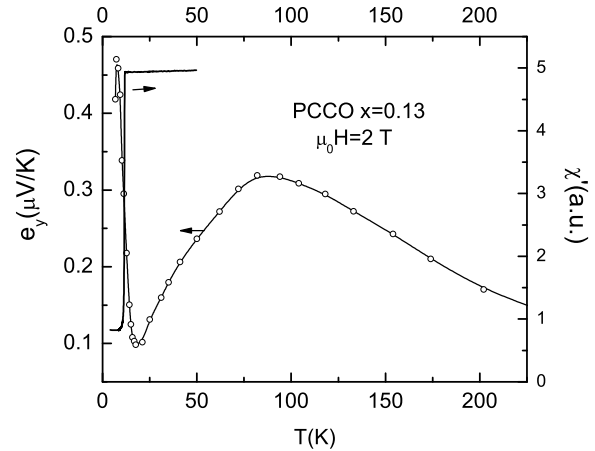}
\end{center}
\caption{Temperature dependence of the Nernst signal in an
underdoped PCCO with \textit{x}=0.13 at $\mu_0$H=2 T. The
H$_{c2}(0)$ of this film is about 7 T and T$_c$ is 11.8 K. Solid
line is the real part of the zero-field AC susceptibility.}
\label{Fig5}
\end{figure}

Before showing all the normal state Nernst effect data, let us
first compare the vortex Nernst signal (T$<$T$_c$) and the normal
state (T$>$T$_c$) for H=2 T in an underdoped \textit{x}=0.13 film.
As shown in Fig.~\ref{Fig5}, two peaks are prominent in the
temperature dependence of the Nernst signal. The lower temperature
peak is produced by vortex motion in the superconducting state and
the higher temperature peak is from the normal state
quasi-particles. In strong contrast to the Nernst effect in
hole-doped cuprates,~\cite{WangYY2} the normal state Nernst signal
is much larger and its magnitude is comparable to the vortex
Nernst signal. As the field approaches H$_{c2}$, the vortex Nernst
signal decrease quickly, but the normal state signal increases
linearly, as seen in Fig.~\ref{Fig1}. In the following, the normal
state Nernst signal is taken at H=9 T, which is greater than
H$_{c2}(0)$ for all the PCCO films.

\begin{figure}
\begin{center}
\includegraphics[scale=1.3]{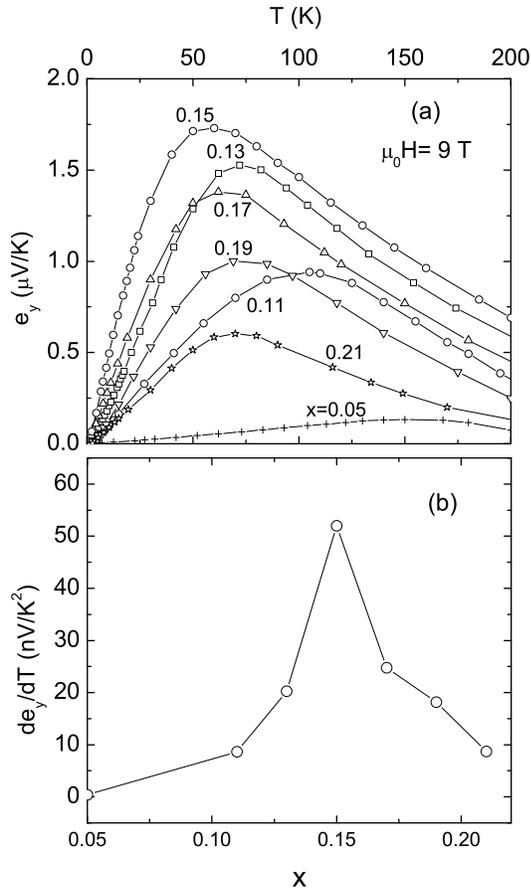}
\end{center}
\caption{(a) Temperature dependence of normal state Nernst signal
at $\mu_0$H=9 T for all the doped PCCO films, (b) Doping
dependence of the initial slope of the Nernst signal curves in (a)
at low temperatures.} \label{Fig6}
\end{figure}

Although our focus of interest is the Nernst effect at the doping
extremes, we measured the normal state Nernst effect on PCCO films
systematically throughout the entire doping range, from the
extremely underdoped (\textit{x}=0.05) to the highly overdoped
(\emph{x}=0.21). Fig.~\ref{Fig6} shows the temperature dependence
of the normal state Nernst signal $e_y$ for all the doped PCCO
films. The Nernst signal increases as temperature decreases,
reaches a peak at a certain temperature then decreases linearly
towards zero as T$\rightarrow$0. The magnitude of the Nernst
signal and its temperature dependence are rather similar for all
the dopings except the extremely underdoped \textit{x}=0.05. The
linear temperature dependence of the the Nernst signal below the
peak temperatures is as predicted for quasi-particles (see later
discussion). In Fig.~\ref{Fig6}(b), the doping dependence of the
slope ($de_y/dT$) of the low temperature Nernst signal is shown.
We see that $de_y/dT$ has a maximum at the optimal doping and
decreases rapidly with underdoping and overdoping.

\subsection{Discussion}

\begin{center}
\textit{One-band transport}
\end{center}

The anomalously large normal state Nernst effect that we observed
throughout almost the entire doping range strongly contrasts with
the hole-doped cuprates and with normal metals in which the
magnitude of the Nernst signal is in the order of $nV/K$. To
understand this, we attempted a simple comparison with
conventional theories for a one-band (single-carrier) system.
Within Boltzmann theory, the Nernst coefficient can be expressed
as~\cite{Vadim},
\begin{equation}\label{2}
\nu=\frac{\pi}{3}\frac{k_{B}^2T}{eB}\frac{\partial{tan\theta_H}}{\partial{\epsilon}}=\frac{\pi^{2}k_{B}^{2}T}{3m^*}\frac{\partial{\tau}}{\partial{\epsilon}}|_{\epsilon_F}
\end{equation}
where $tan\theta_H$ is the Hall angle, $m^*$ is the electron mass
and $\tau$ is the scattering time. Eq.~\ref{2} shows that the
Nernst signal is linearly dependent on the temperature and our low
temperature data at all dopings is consistent with this
prediction. To estimate the magnitude of the low temperature
Nernst signal, we can replace
$\frac{\partial{\tau}}{\partial{\epsilon_F}}$ with
$\frac{\tau}{\epsilon_F}$ by assumption of a weak energy
dependence of $\tau$ at the Fermi energy $\epsilon_F$. This
gives~\cite{Behnia}
\begin{equation}\label{3}
e_y=\nu B=283\omega_c \tau \frac{k_BT}{\epsilon_F}\mu V/K,
\end{equation}
where $\omega_c=eB/m^*$ is the cyclotron frequency. Eq.~\ref{3}
suggests that the Nernst signal is proportional to $\omega_c \tau$
and inversely proportional to Fermi energy.

For a single carrier system, $\omega_c\tau$ can be estimated from
the residual resistivity $\rho_0$ and the carrier density $n$,
i.e., $\omega_c\tau=B/\rho_0 ne$ (which is equivalent to
$\omega_c\tau=tan\theta_H$, here $\theta_H$ is the Hall angle).
For the optimally-doped \textit{x}=0.15, normal state resistivity
measurement gives $\rho_0$=57 $\mu \Omega cm$. The estimate of the
carrier density $n$ is rather unclear. For comparison, we estimate
$n$ in two different ways. From the Hall coefficient
$R_H(0)$,~\cite{YoramQCP} one gets
$n_{Hall}=\frac{1}{R_H(0)e}$=$4.2\times 10^{21}/cm^3$ and then
$\omega_c\tau$=0.022 at 9 T. Another way is by assuming 0.15
electron/unit cell,
$n_{cell}$=$\frac{0.15}{3.95^2\times6}$\AA$^{-3}$=1.58$\times10^{21}/cm^3$,
which yields $\omega_c\tau$=0.059. The Fermi energy $\epsilon_F$
can be obtained from ARPES~\cite{Armitage, Sato} where
$\epsilon_F\sim$0.61 eV (using the Fermi wave number
$k_F\sim0.7\pi/a$ and $\epsilon_F=\frac{\hbar^2k^2_F}{2m^*}$) and
thus $\epsilon_F/k_B\approx$7100 K. Inserting these numbers into
the expression for $e_y$ in Eq.~\ref{3}, we find $de_y/dT$=1.0
nV/K$^2$ (from $n_{Hall}$) and 1.8 nV/K$^2$ (from $n_{cell}$).
These values are more than one order of magnitude smaller than the
measured value of 53 nV/K$^2$ [see Fig~\ref{Fig6}(b)].

Now we apply this estimation to the overdoped sample. The numbers
for $x$=0.19 are $\rho_0$=20 $\mu \Omega cm$,
$n_{Hall}$=$6.5\times 10^{21}/cm^3$ and $n_{cell}$=$2.01\times
10^{21}/cm^3$; $\omega_c\tau$ then is 0.04 and 0.139 respectively
for H=9 T. Although the Fermi energy of $x$=0.19 is still unknown,
we find from APRES experiments~\cite{Matsui2} that $k_F\sim
0.7\pi/a$ remains nearly the same for $x$ from 0.15 to 0.17.
Therefore, we expect that $\epsilon_F$ should also be
approximately 7100 K at $x$=0.19. Plugging these numbers into
Eq.~\ref{3}, we get $de_y/dT$=1.6 nV/K$^2$ and 5.6 nV/K$^2$ for
$n_{Hall}$ and $n_{cell}$ respectively. If we estimate the carrier
density from the area of the Fermi pockets of \textit{x}=0.19
obtained from ARPES~\cite{Matsui2} and the spin-density-wave (SDW)
calculation,~\cite{Lin} we get $n_{FS}=4.3\times10^{21}/cm^3$.
This yields $de_y/dT \sim$ 2.5 nV/K$^2$. All these estimated
values are smaller than the measured value of 20 nV/K$^2$ for
$x$=0.19.

The large difference of these simple estimates from our
experiments shows that a conventional one-band model is not
applicable to PCCO, even in the highly overdoped region with a
large hole-like FS. This suggests that there may be an unusual
energy dependent scattering at the FS which enhances Eq.~\ref{2}
over our simple assumption above.

\begin{center}
\textit{Two-band transport}
\end{center}

The sign change and the temperature dependence of both Hall
coefficient~\cite{YoramQCP} and thermopower,~\cite{PCCOTEP} the
anomalously large Nernst signal and magnetoresistance (to be shown
next) can not be explained by a one-band model. Therefore, one has
to consider a two-carrier transport model for the electron-doped
cuprates.

We start with an estimation of the Nernst signal within a two-band
Boltzmann framework. In this model, assuming an identical
relaxation for both electron and hole carriers,
Oganensyan~\cite{Vadim} found that the Nernst signal is maximal
when the bands are exactly compensated ($n_h=n_e$) and the
expression of the Nernst signal becomes
\begin{equation}\label{5}
e_{y}=B\nu=\frac{2\pi^{2}}{3}\frac{k_{B}^{2}T\tau}{e\hbar}\frac{1}{(k_{F}\ell_{B})^{2}}
\end{equation}
Inserting $k_F \sim 0.5$ \AA$^{-1}$ obtained from ARPES
experiments~\cite{Armitage, Sato} and $\tau\sim 4.16\times
10^{-13}$ s from optics ($1/\tau=80$ cm$^{-1}$ for T just above
T$_c$)~\cite{Homes} for PCCO \textit{x}=0.15, we get $de_y/dT \sim
$25 nV/K$^2$ for $\mu_0$H=9 T. This simple estimation is slightly
smaller than the measured value of 53 $\mu V/K^2$, but suggests
that this two-band Boltzmann model could explain the Nernst effect
in PCCO.

The enhanced Nernst effect in a two-band system can be easily
understood from the expressions for the Nernst signal and
thermopower $S$ (Ref.~\cite{Bel}),
\begin{eqnarray}
e_{y}&=&S(\frac{\alpha^{h}_{xy}+\alpha^{e}_{xy}}{\alpha^{h}_{xx}+\alpha^{e}_{xx}}-\frac{\sigma^{h}_{xy}+\sigma^{e}_{xy}}{\sigma^{h}_{xx}+\sigma^{e}_{xx}})\label{6}\\
&=&S(tan \theta_T-tan \theta_H)\label{7}
\end{eqnarray}
and
\begin{equation}\label{8}
S=\frac{\alpha^{h}_{xx}+\alpha^{e}_{xx}}{\sigma^{h}_{xx}+\sigma^{e}_{xx}}
\end{equation}
here $h$ and $e$ denote the hole and electron bands. Considering
the charge carrier symmetry, $\sigma^{h}_{xy}=-\sigma^{e}_{xy}$
and $\alpha^{h}_{xy}=\alpha^{e}_{xy}$, the second term in
Eq.~\ref{6} will vanish and the first term will be enhanced.
Meanwhile, the counterflow of carriers with different sign in
Eq.~\ref{7}, i.e., $\alpha^h_{xx}=-\alpha^e_{xx}$, will lead to a
small thermopower. Thus, in PCCO, the enhanced Nernst signal and
small thermopower near optimal doping supports the two-band model.
For the underdoped \textit{x}=0.13 and 0.11, a larger thermopower
is found~\cite{PCCOTEP} and the Nernst signal is still large. This
could be due to a contribution from the emerging hole band, as
found in the ARPES experiments.~\cite{Armitage, Matsui2}
\begin{figure}
\begin{center}
\includegraphics[scale=0.8]{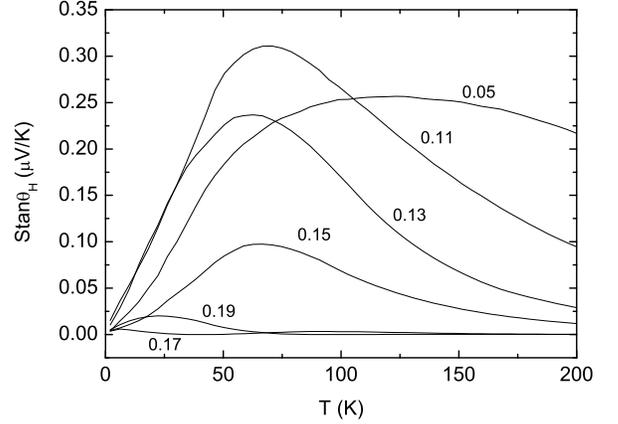}
\end{center}
\caption{Temperature dependence of $Stan\theta_H$ of PCCO films at
 $\mu_0$H=9 T.} \label{Fig7}
\end{figure}

The small thermopower~\cite{PCCOTEP} and sizable Nernst signal
(comparable to the optimal doping) in the overdoped
\textit{x}=0.19 also suggests a possible two-band contribution.
However, this appears to be incompatible with the single large
hole pocket seen in ARPES.~\cite{Matsui2} To explore this further,
we look at $Stan\theta_H$ obtained from the
thermopower~\cite{PCCOTEP} and the Hall angle
measurements,~\cite{YoramHoleSC} for all the doped PCCO films. As
seen in Fig.~\ref{Fig7}, $Stan\theta_H$ decreases with increasing
doping and its magnitude is much smaller than the Nernst signal in
Fig.~\ref{Fig6} for all the dopings except the extremely
underdoped \textit{x}=0.05. As we just described, to have a large
Nernst signal, the difference between $Stan\theta_H$ (the second
term in Eq.~\ref{7}) and $Stan\theta_T$ (the first term in
Eq.~\ref{7}) has to be large. In PCCO, the large difference in the
magnitude of $e_y$ and $Stan\theta_H$ (see Fig.~\ref{Fig6} and
Fig.~\ref{Fig7}) suggests a large thermal Hall angle $\theta_T$
(recall that \textit{S} is small). Therefore, the large value of
$tan \theta_T$ indicates a two-carrier contribution. This is also
seen in the overdoped PCCO from Fig.~\ref{Fig6} and
Fig.~\ref{Fig7}, strongly suggesting the incompatibility with the
single hole-like FS expectation. Note, for a single carrier
system, such as the extremely underdoped $x$=0.05 PCCO, the small
Nernst signal is comparable to $Stan\theta_H$, suggests a
negligible $Stan\theta_T$. It is worth mentioning that two-band
transport in the highly overdoped PCCO is also consistent with the
high-field nonlinear Hall resistivity.~\cite{HighFieldHall}

We have shown that the anomalous large normal state Nernst effect
and small $Stan\theta_H$ in the overdoped PCCO films are
qualitatively consistent with a two-band model. However, this is
contrary to the ARPES experiments,~\cite{Matsui2} in which a
simple hole-like Fermi pocket is found for doping $x\geq$0.17.
Another possible reason for the enhanced Nernst signal, as
indicated in Eq.~\ref{2}, could be a strong anisotropic energy
dependence of the scattering $\tau(\epsilon_F)$ at the Fermi
surface. The remanent contribution of ``hot spots'' near the
($\pi/2, \pi/2$) region to the Nernst signal could be dominant
even in the overdoped regime.

\begin{figure}
\begin{center}
\includegraphics[scale=0.8]{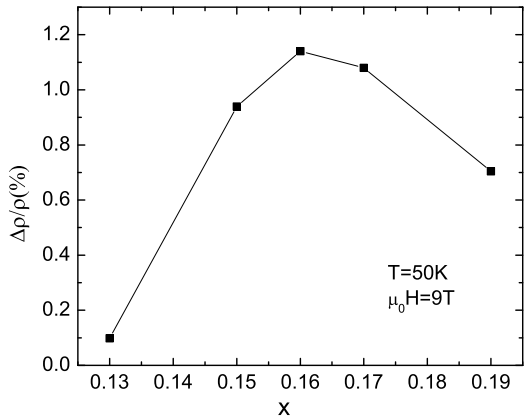}
\end{center}
\caption{Doping dependence of the normal state magnetoresistance
in PCCO films at T=50 K (T is set at 50 K in order to avoid the
low temperature superconducting fluctuations).} \label{Fig8}
\end{figure}

The magnetoresistance in a two-carrier system is closely related
to the Nernst effect. From Eq.~\ref{6} and \ref{8}, one can
rewrite the Nernst coefficient as~\cite{Belthesis}:
\begin{equation}\label{9}
\nu=\frac{\nu_h\sigma_h+\nu_e\sigma_e}{\sigma}+\frac{\sigma_h
\sigma_e(S_h-S_e)(\sigma_hR_h-\sigma_eR_e)}{\sigma^2}
\end{equation}
The magnetoresistance for a two-band system is~\cite{Fournier}
\begin{equation}\label{10}
\frac{\triangle\rho}{\rho}=\frac{\sigma_h \sigma_e(\sigma_h
R_h-\sigma_e R_e)^2B^2}{\sigma^2}
\end{equation}
The second term of Eq.~\ref{9} is responsible for the potentially
larger Nernst signal with respect to a single carrier system. The
factor $(\sigma_hR_h-\sigma_eR_e)=(\mu_h-\mu_e)$ can reach a
maximum value if the mobilities are large and $\mu_h=-\mu_e$,
which will lead to an enhanced Nernst signal. From Eq.~\ref{10},
we can identify the same mobility coefficient,
$(\sigma_hR_h-\sigma_eR_e)$, found in the Nernst coefficient. This
indicates that a maximum of the magnetoresistance is likely to
coincide with a maximum of the Nernst coefficient as the doping
and the mobilities change.~\cite{Jiang, Fournier} In
Fig.~\ref{Fig8}, we show the magnetoresistance at 50 K as a
function of Ce content. We see that the transverse
magnetoresistance is large and positive (compared to the
magnetoresistance in the p-doped cuprates, which is one order of
magnitude smaller) for all the superconducting films. The maximum
magnetoresistance occurs around optimal doping, at which doping,
the Nernst signal [Fig.~\ref{Fig6}(b)] also reach a maximum. The
strong correlation between Nernst effect and the magnetoresistance
strongly suggests that PCCO is a two-band system for dopings in
the superconducting dome region of the phase diagram.

Finally, we discuss the higher temperature peak feature that found
in the temperature dependence of the Nernst signal
[Fig.~\ref{Fig5}(a)]. A similar peak feature was found in
thermopower measurements which suggests a common origin. A prior
work~\cite{Tallon} proposed that the thermoelectric power of the
underdoped hole-doped cuprates could reveal the opening of a
pseudogap. The pseudogap increases the thermopower leading to a
broad maximum above T$_c$ but below T$^*$. For the PCCO system, we
also observed a maximum thermopower at fairly high temperature
(between 50 K and 100 K) with respect to T$_c$ in the underdoped
region. It is possible that the thermopower is influenced by the
high temperature gap found in the optics and ARPES
measurements.~\cite{Zimmers} The transverse thermoelectric effect,
i.e., Nernst effect, also presents a broad higher temperature peak
in the entire doping range. The peak feature could also be a
result of the influence of the high energy gap and fluctuations in
the overdoped region. Further understanding of this will require
future study.

The peak feature in the temperature dependence of the Nernst
signal could also be a result of a phonon-drag effect. As proposed
by Behnia \textit{et al.}~\cite{Behnia} for Bismuth, in the
Ettingshausen geometry (transverse temperature gradient generated
by external magnetic field when a longitudinal electrical current
is present), when electrons interact with phonons, the electric
current gives rise to an entropy current of phononic origin, and a
significant Ettingshausen effect. Since the Onsager relation ties
the amplitudes of the Ettingshausen and Nernst effects, this
implies that the Nernst effect should also be enhanced. However,
the application of this model to the PCCO case is not clear at
this time and future work is required.

\subsection{Conclusion}

We performed measurements of the magnetic field driven normal
state Nernst effect on electron-doped cuprate
Pr$_{2-x}$Ce$_x$CuO$_{4-\delta}$ films over a wide range of doping
and temperature. We find an anomalously large Nernst signal near
optimal doping, which is consistent with prior reports. More
interestingly, the Nernst signal is still large in the highly
overdoped films and the slightly underdoped films. This can not be
explained by a single-band model, and a two-band model has to be
considered. The qualitative consistence between the experimental
transport data and the two-band model for the overdoped films,
which appears inconsistent with ARPES experiments, will require
further research.

\begin{acknowledgements}
We thank A. Millis, K. Behnia, V. Yakovenko, W. Yu, V. Oganesyan
and H. Kontani for fruitful discussions. P.L. thanks H. Balci for
the Nernst effect measurements. We acknowledge the support of NSF
under Grant DMR-0352735.
\end{acknowledgements}

\end{document}